# Protostar mass functions in young clusters

**Philip C. Myers**

Harvard-Smithsonian Center for Astrophysics, 60 Garden Street,

Cambridge MA 02138 USA

pmyers@cfa.harvard.edu

**Abstract.** In an improved model of protostar mass functions (PMFs), protostars gain mass from isothermal cores in turbulent clumps. Their mass accretion rate is similar to Shu accretion at low mass, and to reduced Bondi accretion at high mass. Accretion durations follow a simple expression in which higher-mass protostars accrete for longer times. These times are set by ejections, stellar feedback, and gravitational competition, which terminate accretion and reduce its efficiency. The mass scale is the mass of a critically stable isothermal core. In steady state, the PMF approaches a power law at high mass due to competition between clump accretion and accretion stopping. The power law exponent is the ratio of the time scales of accretion and accretion stopping. The luminosity function (PLF) peaks near 1 $L_{\odot}$, due to inefficient accretion of core gas. Models fit observed PLFs in four large embedded clusters. These indicate that their underlying PMFs may be top-heavy compared to the IMF, depending on the model of protostar radius.

## 1. Introduction

The mass function of stars at birth, or the initial mass function (IMF), is a fundamental property of stars, yet its origin remains a major unsolved problem in astrophysics. The IMF has similar properties of shape, mass scale, and high-mass slope among field stars, open clusters, and young clusters (Kroupa 2002, Chabrier 2005, Bastian et al. 2010). Many explanations for the origin of the IMF have been proposed (Bonnell et al. 2001; Padoan & Nordlund 2002; Pudritz 2010; Parravano et al. 2011; Hennebelle & Chabrier 2011, hereafter HC11; Hopkins 2012).

In the most widely-discussed explanation, an IMF distribution of masses arises as a cluster-forming clump fragments into condensations, or cores, due to self-gravity and turbulence. In such turbulent fragmentation models, the mass distribution of cores (CMF) is a mass-shifted version of the IMF (Padoan & Nordlund 2002, Hennebelle & Chabrier 2008, Hennebelle & Chabrier 2009, Hopkins 2012), matching observational studies of cores in nearby star-forming regions (Motte et al. 1998, Alves et al. 2007, Könyves et al. 2010).

Whether the CMF generates the IMF by a simple mass scaling has been questioned, because core properties alone do not necessarily set the accretion history of a protostar (Clark et al. 2007, Smith et al. 2009). Ejections, gravitational competition and stellar feedback may act to stop accretion onto a protostar before all the available core gas is accreted (Reipurth & Clarke 2001, Bonnell et al. 1997, Velusamy & Langer 1998). Furthermore, cores in star-forming regions are generally not isolated from their surrounding clump gas (Teixeira et al. 2005), allowing massive stars to accrete from beyond the core (Smith et al. 2009, Wang et al. 2010).

The dependence of protostar mass on accretion history motivates mass function models based on "stopped accretion" (HC11). In these models, the duration of accretion is more important than the initial core mass in setting the final protostar mass (Basu & Jones 2004, Myers 2009b, 2012; hereafter Paper 1). This property is indicated by some simulations of competitive accretion, which show that the final mass depends more strongly on accretion duration than on mean accretion rate (Bate & Bonnell 2005, Bate 2012). In such stopped accretion models, the mass distribution of still-accreting protostars (PMF) is distinct from that of pre-main sequence stars, which have stopped accreting (McKee & Offner 2010, Paper 1). This paper describes the PMF and the corresponding luminosity function in embedded clusters.

Most stopped accretion models assume that accretion is equally likely to stop in any infinitesimal time interval. Then the probability that accretion lasts until $t$ and stops between $t$ and $t + dt$ is a declining exponential function of $t$ (Basu & Jones 2004, Bate & Bonnell 2005). For such models, the competition between accretion and accretion stopping yields mass functions whose shape resembles that of the IMF (Myers 2009b). However, these models remain incomplete until their parameters have a clear physical basis (HC11).

The model presented here is an improved version of those in previous papers (Myers 2011, Paper 1). In this paper the mass scale and initial accretion rate are based on the physics of thermal, self-gravitating gas and are not simply fitting parameters. The accretion model has two parameters rather than three, and it leads to a simpler analytic expression for the protostar mass as a function of time. The model fits luminosity functions in young clusters, and then infers the underlying cluster mass functions and initial properties of their star-forming gas.

In this paper, Section 2 justifies the thermal and spherical approximations used in Section 3, Section 3 formulates the model, Section 4 compares results and observations, Section 5 discusses the paper, and Section 6 gives a summary. The variables and parameters used in the paper are summarized in Table 1.

## 2. Thermal and spherical core models

This section summarizes observed properties of cores and clumps, and justifies the thermal, spherical description of star-forming cores used in the model of Section 3.

It is generally accepted that low-mass stars are born in "dense cores," or condensations whose internal velocity dispersion is dominated by thermal motions and whose size and mean density are of order 0.05 pc and $10^4$ cm$^{-3}$, according to observations of spectral lines of $NH_3$ and other tracers (Myers & Benson 1983, Beichman et al. 1986). The "clump" gas surrounding such cores has filamentary structure, with spatial extent and mean density of order 0.5 pc and $10^3$ cm$^{-3}$ (Hatchell et al. 2005, Enoch et al. 2006). Its velocity dispersion is dominated by supersonic motions, according to observations of lines of $^{13}CO$ and other tracers (Bergin & Tafalla 2007). The motions and structure of clump gas resemble predictions of simulations of HD and MHD turbulence (MacLow & Klessen 2004).

In regions of intermediate and high-mass star formation, starless $NH_3$ cores are more massive, and typically have trans-sonic or moderately supersonic line widths ~ 1 km s$^{-1}$ (Olmi et al. 2010, Pillai et al. 2011, Sánchez-Monge 2013). Nonetheless these more massive cores resemble low-mass cores in that they are significantly less turbulent than their surrounding gas. A significant fraction (7/23) of starless cores in embedded clusters with protostellar luminosity > 300 $L_{\odot}$ have thermally dominated $NH_3$ line widths (Sánchez-Monge 2013). In cores in the Spokes cluster, supersonic line widths decrease toward sonic values as the critical density of the tracer line increases (Pineda & Teixeira 2013).

Cores and clumps have complex structure. Clumps which harbor embedded clusters resemble filamentary networks, some of whose filaments meet in nodes and hubs having relatively high density of gas and stars (Molinari et al. 2010, André et al. 2010, Peretto et al. 2013). In turn, filaments harbor cores, and in some cases cores have regular spacing of a few core diameters (Schneider & Elmegreen 1979, Schmalzl et al. 2010). The internal structure of most cores is elongated, with aspect ratio ~2 at resolution of a few 0.01 pc (Myers et al. 1991, Ryden 1996). In interferometric studies with finer resolution the aspect ratio is significantly greater (Pineda et al. 2010, Bourke et al. 2012). Envelopes around some accreting protostars appear filamentary, suggesting that the accretion flow itself may be filamentary (Lee et al. 2012).

Despite these nonspherical features of core structure, it has proven useful to model starless cores as self-gravitating isothermal spheres in equilibrium (Shu 1977, Alves et al. 2001, McKee & Tan 2003, Fatuzzo et al. 2004, Tafalla et al. 2004, Keto & Caselli 2008). Circularly averaged column density maps match models of self-gravitating isothermal spheres in B68 (Alves et al. 2001), in L1498 and L1517B (Tafalla et al. 2004), and in ten Bok globules. More than half of these globules were found to be close to the point of critical stability (Kandori et al. 2005). Cores in the Pipe Nebula have been modeled as equilibrium isothermal spheres bounded by the pressure of the surrounding gas. In the Pipe Nebula, the gas which provides the external pressure has supersonic velocity dispersion, as required to support the weight of the cloud (Lada et al. 2008).

In this paper, the typical mass is that of a critically stable isothermal sphere (CIS), or Bonnor-Ebert sphere (Bonnor 1956; Ebert 1955), a useful description of a self-gravitating isothermal condensation on the verge of star-forming collapse. The effects of rotation, magnetic fields, and nonisothermal equations of state are beyond the scope of this paper, but they should be included in a fuller treatment.

## 3. Model of accretion and accretion stopping

### 3.1. Accretion model

In this paper "accretion" refers to protostar mass gain from its environment. For most of the accretion history, this mass gain is dominated by accretion through the circumstellar disk. It is assumed that the rate of mass gain from the disk to the protostar is equal to the rate of mass gain from the core-clump environment to the disk, as expected in steady state, and for simplicity both processes are here called accretion.

The accreting protostar of mass $m$ is assumed to gain mass from its surrounding core and clump at a rate $\dot{m}$, according to

$$\dot{m} = \dot{m}_0[1 + (m/m_0)^2]^{1/2} \tag{1}$$

where the initial mass accretion rate $\dot{m}_0$ and the mass scale $m_0$ are parameters independent of mass, time, and position in the cluster. This independence of cluster position neglects the likely increase of accretion rate near the clump center of mass, due to global infall of clump gas (Smith et al. 2009, Wang et al. 2010).

Equation (1) is based on the assumption that all stars begin their accretion from centrally condensed, thermally supported "core" gas. The surrounding turbulent clump gas provides mass for accretion and exerts pressure on the core. Such a two-component form of the accretion rate is needed to account for both low-mass and massive star formation. In the version adopted here, denoted 2CA, the constant term represents low-mass star formation from an isothermal sphere (Shu 1977). The mass-dependent term represents massive star formation, as in competitive accretion models (Bonnell et al 2001a,b). Equation (1) resembles expressions in McKee & Tan (2003) (MT), McKee & Offner 2010 (MO), OM, and in Paper 1. However, here the accretion rate depends on current protostar mass, and not on final mass as in MO and OM.

In the high-mass limit of equation (1), the accretion rate of turbulent clump gas increases as a power of the protostar mass. The value of this power is set by the stopping probability and by the requirement that the mass function approach a declining power law at high mass, as in the IMF. When the stopping probability is equal in each time interval, the high-mass accretion rate varies as $\dot{m} \sim m$. This accretion rate dependence on mass is weaker than the rate $\dot{m} \sim m^2$ for stationary accretion from a uniform medium (Bondi 1952).

In this model thermal core gas sets the initial accretion rate of a protostar, but not its final mass. The final mass $m_f$ of a protostar may originate from a subregion of the parent core, as when a member of a small group is ejected (Reipurth & Clarke 2001). Or it may originate from a region larger than the core, as when a massive star draws from core gas and then from clump gas (Smith et al. 2009, Wang et al. 2010). The final mass $m_f$ is assumed to come from an “original mass” $M$, which is independent of the core mass.

The mass flow from $M$ to $m_f$ is assumed inefficient, due to the processes which limit accretion. These include protostar ejection (Reipurth & Clarke 2001), gravitational competition (Bonnell et al. 1997), gas dispersal by stellar feedback due to outflows (Velusamy & Langer 1998), radiative heating and ionization (Offner et al. 2009, Klassen et al. 2012), and exhaustion of the original gas supply. Studies of the CMF and the IMF indicate that the typical star mass is less than the typical core mass by a factor ~ 3, which applies on average but not for each individual star (Rathborne et al. 2009, Michel et al. 2011). It is assumed here that $m_f = \varepsilon M$ with $\varepsilon = 1/3$ (Alves et al. 2007, Matzner & McKee 2000).

The protostar mass accretion rate is similarly inefficient, due to dispersal of some of the original mass before it accretes onto the disk, or before it accretes from the disk onto the protostar. An analysis of the TMC-1 system indicates an accretion rate efficiency of 0.25 with respect to spherically symmetric SIS collapse (Terebey et al. 2006). It is assumed here that the efficiencies of accretion and of accretion rate are equal,

$$\dot{m} = \varepsilon \dot{M} . \qquad \textbf{(2)}$$

Here $\dot{M}$ is the accretion rate from the original mass $M$ onto the protostar-disk system in the absence of any dispersal.

The assumption of constant efficiency of mass and mass accretion rate is a highly simplified average over accretion processes which are more complex in space and time, according to high-resolution observations (Lee et al. 2012) and simulations (Krumholz et al. 2012, Girichidis et al. 2011). Also, this assumption neglects neglect temporal inefficiency due to episodic or nonsteady accretion (Dunham & Vorobyov 2012), to be discussed in Section 4.1.

### 3.2. Mass scale

The original mass $M_0$ of the protostar mass scale $m_0$ is assumed to equal the mass of a Bonnor-Ebert sphere (Bonnor 1956; Ebert 1955), also known as a critically stable isothermal sphere (CIS), discussed in Section 2. The protostar mass scale $m_0$ reflects the physics of a critically stable core, which sets $M_0$, and it reflects the physics of gas dispersal during infall and disk accretion, which set $\varepsilon$.

The mass scale is chosen to be $M_{\rm CIS}$, rather than $M_{\rm SIS}$, the mass of a truncated isothermal sphere in the limit of infinite central density (Chandrasekhar 1939). The CIS is more appropriate than the SIS as a mass scale for thermal star formation. The CIS is critically stable, and is therefore on the verge of star-forming collapse. In contrast the SIS is infinitely unstable, and is less likely to be achieved as an initial state (Whitworth et al. 1996).

The SIS is nonetheless useful for this paper, because of a simple relation between a CIS and a truncated SIS having the same temperature and mean density,

$$M_{CIS} = \Gamma_{IS} M_{SIS} \quad , \tag{3}$$

where the constant factor $\Gamma_{\rm IS}$ is defined by

$$\Gamma_{IS} \equiv \left(\frac{\nu_c}{2}\right)^{3/2} \quad . \tag{4}$$

Here $\nu_{\rm c}$ is the negative log-log slope of the density profile of an isothermal sphere at its radius of critical stability (Chandrasekhar 1939). At this radius, $\nu_{\rm c} = 2.434$ (Chandrasekhar & Wares 1949), so that equation (4) gives $\Gamma_{\rm IS} = 1.342$. This factor is close to the IMF slope $\Gamma_{\rm S} = 1.35$ (Salpeter 1955), to be discussed in Section 3.6.

### 3.3. Protostar accretion history

The protostar mass is obtained as a function of its accretion duration $a$, or the time since the start of its accretion, from equations (1) and (2) as

$$m = m_0 \sinh\left(\frac{\dot{m}_0 a}{m_0}\right) \quad . \qquad (5)$$

Equation (5) indicates that the protostar mass increases linearly at early times, as expected from the collapse of a centrally condensed isothermal core. At late times, the mass increases exponentially as it accretes clump gas, reaching massive-star values if its accretion does not stop first.

### 3.4. Accretion stopping

Protostar accretion can stop due to the same processes which limit its efficiency. As noted above in Section 3.1, these include protostar ejection, gravitational competition, gas dispersal by stellar feedback due to outflows, radiative heating and ionization, and exhaustion of the original gas supply. To date, no detailed study of the combination of these processes is available. Here, their combined effect is described in a simple statistical way, as in Fletcher & Stahler (1994) and Paper 1.

In this model a cluster forms protostars at constant birthrate $b$ from time $t = 0$ until the present time $t$. The probability density that a protostar accretes for duration $a$ and stops accreting between $a$ and $a + da$ is

$$p(a) = \frac{\exp(-a/t_s)}{t_s[1 - \exp(-t/t_s)]} \qquad \mathbf{(6)}$$

where the durations $a$ span the range $0 \le a \le t$ and where $t_s$ is the accretion stopping time scale (Paper 1).

When $t >> t_s$ the protostar population approaches a steady state, where the rate of accretion stopping equals the birthrate, and where the number of protostars approaches $bt_s$ (Fletcher & Stahler 1994). Then the probability density $p(a)$ approaches the condition of equally likely stopping and the accretion stopping time scale $t_s$ approaches the mean duration of accretion (Basu & Jones 2004, Paper 1). This steady state condition $t >> t_s$ is satisfied in most embedded clusters, whose oldest stars have age ~ 1 Myr (Gutermuth et al. 2008), much greater than the accretion stopping time scale, ~ 0.1 Myr (Dunham & Vorobyov 2012).

In such a steady state cluster, the mass function of still-accreting protostars has the same shape as the final mass function of the cluster, but with a smaller amplitude and a smaller maximum mass (Paper 1). Assuming that the final mass function matches the IMF, this property allows the IMF to constrain model parameters, as in the next section.

### 3.5. Protostar mass function

The protostar mass function $dN/d\log m$ in steady state is obtained by assuming that the duration of accretion is the most important factor in setting the protostar mass, as found in simulations of competitive accretion (Bate & Bonnell 2005, Bate 2012). Then the probability densities for protostar mass $p(m)$ and accretion duration $p(a)$ are related by $p(m)dm = p(a)da$, and equations (5) and (6) give $p(m)$ as

$$p(m) = \frac{\Gamma(\mu+\nu)^{-\Gamma}}{m_0 \nu} \qquad (7)$$

where $\mu \equiv m/m_0$ is the protostar mass normalized by the mass scale, $\nu \equiv (1+\mu^2)^{1/2}$ is the quadrature sum of 1 and $\mu$, and where the exponent is defined by $\Gamma \equiv m_0/(\dot{m}_0 t_s) = M_0/(\dot{M}_0 t_s)$.

The mass function for $N$ protostars, $dN/d\log m = \ln(10)Nmp(m)$, can be written from equation (7) as

$$\frac{dN}{d\log m} = \ln(10)N\Gamma\left(\frac{\mu}{\nu}\right)(\mu+\nu)^{-\Gamma}. \tag{8}$$

When $\mu << 1$, equation (8) shows that the mass function increases linearly with $\mu$, because at low mass the probability density $p(m)$ in equation (7) approaches a constant value. In contrast, when $\mu >> 1$, the mass function decreases as $\mu^{-\Gamma}$. Thus $\Gamma$ equals the negative of the log-log slope of the mass function at high mass.

The power-law decline of the mass function at high mass is a consequence of the equally likely stopping probability in equation (6), and of the linear mass dependence of the accretion rate at high mass in equation (1). If the mass-dependent clump accretion term were omitted from equation (1), the resulting mass function would decline at high mass as an exponential and not as a power law.

The constancy of the slope $\Gamma$ can be understood in terms of the constancy of its defining parameters. The slope $\Gamma \equiv (m_0/\dot{m}_0)/t_s$ is the ratio of two time scales - the accretion time scale $m_0/\dot{m}_0$ and the accretion stopping time scale $t_s$. This ratio is expected to be constant when both time scales are controlled by the dynamical time of the accreting medium (Basu & Jones 2004). Observational estimates indicate that these time scales are similar, as discussed in Section 3.6.

The 2CA mass function in equation (8) is similar to recently published functions which fit observed estimates of the IMF (De Marchi et al. 2010, Parravano et al. 2011, Myers 2011b, Paper 1). It has the same form as a model of the CMF based on competition between gravitational contraction of a filament and equally likely stopping due to gas dispersal (Myers 2013). On the other hand, the 2CA mass function is more physical than the foregoing expressions, since it is based on gas properties of a thermal core in a turbulent medium.

The 2CA model resembles IMF models of turbulent fragmentation (Padoan & Nordlund 2002, Hennebelle & Chabrier 2008, 2009, Hopkins 2012) in its reliance on thermal and turbulent gas properties. However the 2CA model describes accretion onto protostars and specifies when accretion stops. In contrast, the above turbulent fragmentation models predict the CMF and assume that the CMF is a sufficient initial condition to predict the IMF. The models also differ because the 2CA model is simpler and has fewer parameters.

### 3.6. Initial mass accretion rate

The initial mass accretion rate is related to the high-mass slope of the mass function. From the above definition of $\Gamma$ and the definition of $\Gamma_{\rm IS}$ in equation (3), the initial mass accretion rate can be written

$$\dot{M}_0 = \left(\frac{\Gamma_{IS}}{\Gamma_S}\right)\left(\frac{\Gamma_S}{\Gamma}\right)\left(\frac{M_{SIS}}{t_s}\right). \qquad (9)$$

In equation (9), the factor $\Gamma_{\rm IS}/\Gamma_{\rm S}$ lies within 1% of unity according to equation (4). Thus the high-mass slope $\Gamma$ closely matches the Salpeter value $\Gamma_{\rm S}$ if the initial mass accretion rate matches $M_{SIS}/t_s$, and conversely.

The initial mass accretion rate $\dot{M}_0 = M_{SIS}/t_s$ needed to obtain $\Gamma \approx \Gamma_S$ resembles the accretion rate of a collapsing SIS. This conclusion follows because $M_{SIS}/t_s$ can be written as the accretion rate for pressure-free collapse of a SIS, $M_{SIS}/t_f$, times the ratio of time scales $t_f/t_s$. This ratio appears close to unity, since estimates of the typical duration of accretion $t_s$ and of the free-fall time of star-forming gas $t_f$ are each of order 0.1 Myr in nearby star-forming regions (Evans et al. 2009, Dunham & Vorobyov 2012, Foster et al. 2009, Lada et al. 2010). The typical departure of $t_f/t_s$ from unity is uncertain, but is estimated to be less than a factor of 3.

Thus the initial mass accretion rate is constrained to approximate that of SIS collapse. However this constraint is not specific enough to discriminate among SIS collapse models with and without thermal pressure, with varying initial overdensity, or with varying initial velocity (Shu 1977, Fatuzzo et al. 2004). A detailed model of CIS initial state and SIS accretion rate was presented by Lee et al. (2004).

To make specific predictions, it is assumed that the time scales $t_{\mathrm{f}}$ and $t_{\mathrm{s}}$ are equal, i.e. the mean duration of accretion equals the free fall time of the mass scale gas. Their common value is denoted as the star formation time scale $\tau = t_{\mathrm{f}} = t_{\mathrm{s}}$. Then the mass function slope $\Gamma$ equals the ratio of time scales $\Gamma = (m_0 / \dot{m}_0)/\tau$. This slope matches the Salpeter slope when the initial mass accretion rate onto the protostar is that of inefficient pressure-free collapse of a SIS, $\dot{m}_0 = 8\varepsilon\sigma^3 / (\pi G)$, where $\sigma$ is the thermal velocity dispersion and $G$ is Newton's gravitational constant.

On the other hand, if the time scales $t_{\mathrm{f}}$ and $t_{\mathrm{s}}$ are not equal, the foregoing points indicate that a CIS mass scale and an initial mass accretion rate due to pressure-free SIS collapse give a mass function slope $\Gamma = \Gamma_{\mathrm{IS}}(t_{\mathrm{f}}/t_{\mathrm{s}})$. The slope can thus be "top-heavy" or "bottom-heavy" compared to the IMF depending on the ratio of $t_{\mathrm{f}}$, the free-fall time of mass scale gas and $t_{\mathrm{s}}$, the mean accretion duration.

### 3.7. Characteristic masses

The analytic form of the mass function model allows simple expressions for the mean, median, and modal masses. They are similar to each other, and they reflect the peak but not the width of the broad mass distribution.

The mean mass is obtained from the probability density *p(m)* in equation (7) by integrating *mp(m)* over all masses. The median mass is obtained by setting the cumulative probability to 1/2, and the modal mass is obtained by setting the derivative of the mass function in equation (8) to zero. The mean and median mass become independent of their mass limits when the limits are sufficiently far from the mass scale, i.e. when $m_{\min} << m_0$ and $m_{\max} >> m_0$. Equations (10) and (11) give expressions for these limiting cases. The modal mass, in equation (12), is independent of the limiting values. In each case the mass is given in its dimensionless form, $\mu = m/m_0$:

$$\bar{\mu} = \frac{\Gamma}{\Gamma^2 - 1} \quad , \tag{10}$$

$$\mu_{\text{med}} = \sinh(\ln 2/\Gamma) \quad , \tag{11}$$

$$\mu_{\text{mod}} = \left[\frac{(1+4/\Gamma^2)^{1/2} - 1}{2}\right]^{1/2} \quad . \tag{12}$$

For $\Gamma = 1.342$ as in Section 3.1, equations (10) - (12) give $\bar{\mu} = 1.68$, $\mu_{\text{med}} = 0.540$, and $\mu_{\text{mod}} = 0.630$. These values indicate that the mass scale lies between the median and mean mass, and is greater than the mode of the mass function by the factor 1/0.630 = 1.59. Also, the modal mass is significantly less than the mean mass, $m_{mod}/\overline{m} = 0.375$, since the asymmetric mass function is steep at low mass and shallow at high mass.

For mass limits $m_{min} = 0.01\ M_{\odot}$, $m_{max} = 100\ M_{\odot}$, and for mass scale $m_0 = 0.22\ M_{\odot}$ as adopted below, the mean mass exceeds the limiting value in equation (10) by 1% and the median mass is less than the limiting value in equation (11) by 9%.

## 3.8. Model parameters and comparison with the IMF

The mass scale, initial accretion rate, and stopping probability in equations (1)-(6) require three independent parameters, which can be specified as the thermal gas velocity dispersion $\sigma$; the mass efficiency $\varepsilon$ of star-forming gas, and the star-forming time scale $\tau$. This time scale denotes the mean accretion duration and the free fall time of the mass scale gas, which are assumed equal to each other. In combination with the model equations, these three parameters completely specify the mass function model.

Parameter values which match observational estimates, and which yield a mass function matching standard representations of the IMF are $\sigma = 0.19$ km s$^{-1}$, $\varepsilon = 1/3$, and $\tau =$ 0.12 Myr. These are based on observations of nearby regions of star formation. There $NH_3$ line observations of starless cores in Perseus give median thermal velocity dispersion 0.19 km s$^{-1}$ corresponding to temperature 11 K (Foster et al. 2009), the mass function of starless cores in the Pipe Nebula appears shifted from the IMF by a factor 3 (Alves et al. 2007), and modelling of the evolution of the spectral energy distributions of protostars indicates that the embedded phase has a duration of 0.12 Myr (Dunham & Vorobyov 2012).

These parameter values yield a mass scale $m_0 = 8\varepsilon\sigma^3\tau\Gamma/(\pi G) = 0.22\ M_{\odot}$. Appplying equations (10)-(12) gives median, modal, and mean mass respectively 0.12, 0.14, and 0.37 $M_\odot$. The mean mass agrees closely with the mean mass estimated from observations, 0.36 $M_\odot$ (Weidner & Kroupa 2006).

Variation of parameter values from the above choices affect the mass function mass scale but not its shape. In this model, the mass function shape is constant because its log-log slopes, unity at low mass and -$\Gamma$ at high mass, are independent of parameter values. The mass scale is also unchanged if its parameters vary so that the product $\varepsilon\sigma^3\tau$ is unchanged. Neglecting variation in $\varepsilon$, this condition is equivalent to varying the mean gas density and temperature as $\bar{n} \sim T^3$, i.e. where star-forming gas is hotter, it must also be denser. Observational evidence for small variation in mass scale is discussed in Section 5.1.

The 2CA mass function based on the above equations and parameters agrees well with the IMF estimates of Kroupa (2002) and Chabrier (2005). Each of these segmented IMF estimates is represented in Figure 1 by a smooth function with no discontinuities (Myers 2010). The 2CA mass function lies between the two IMF estimates and has essentially the same high-mass slope and the same mass scale. The low-mass limit is a power law of slope unity, due to the combination of ELS and constant initial mass accretion rate. At low mass the Kroupa and Chabrier estimates diverge, while the 2CA model approximates the geometric mean of the two.

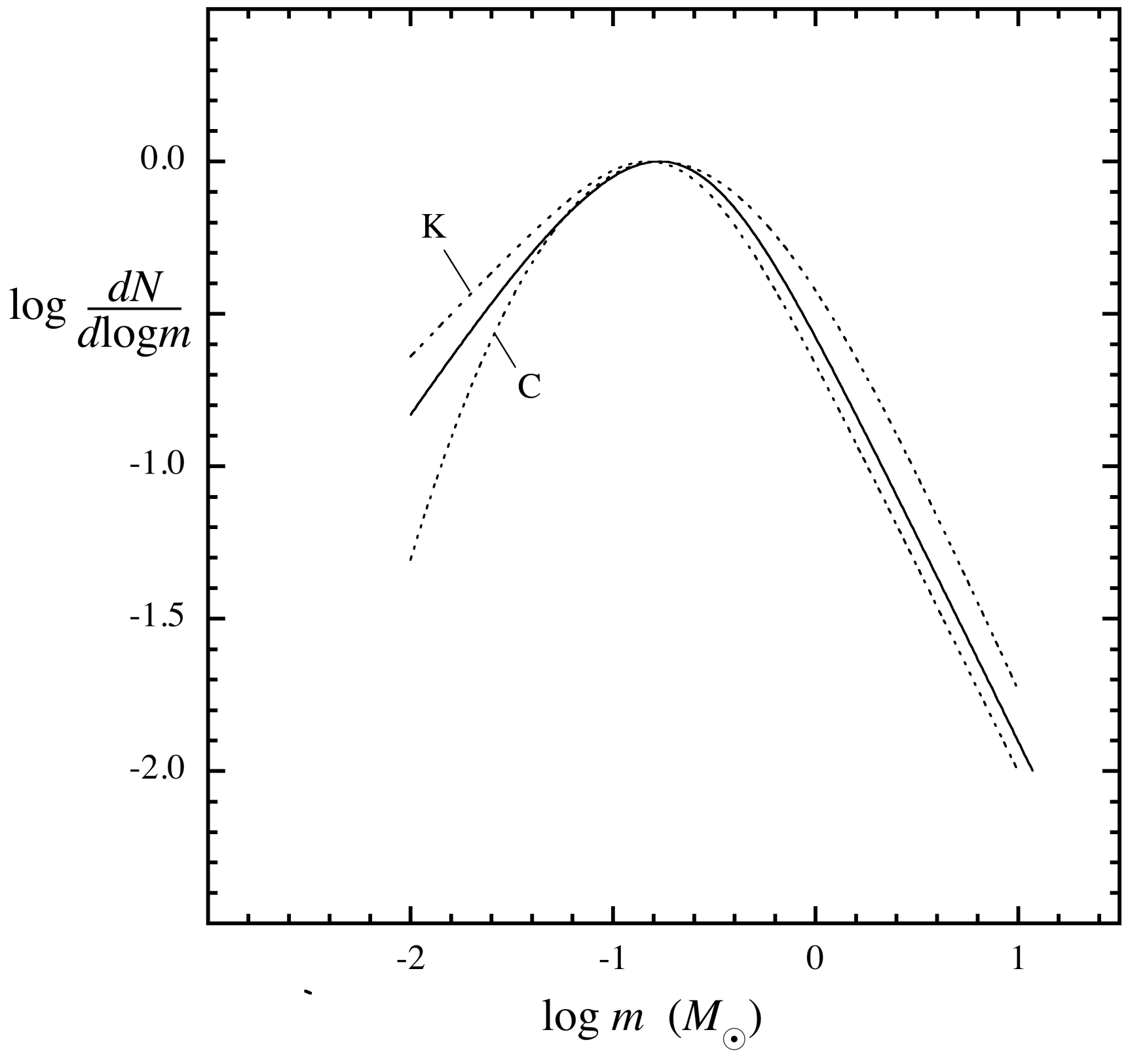

**Figure 1.** Observed and model mass functions. *Solid curve*, mass function based on two-component accretion model and equally likely stopping in a steady-state cluster, with thermal velocity dispersion $\sigma$ = 0.19 km s$^{-1}$, star formation mass efficiency $\varepsilon$ = 1/3, and star formation time scale $\tau$ = 0.12 Myr. *Dotted curves*, representations of the initial mass functions of Kroupa (2002) and Chabrier (2005).

The good agreement between the model protostar mass function and the IMF curves in Figure 1 indicates that the model can reproduce the low-mass shape and the mass scale of the IMF, with parameter choices based on observed properties of star-forming gas. The high-mass slope matches that of the IMF by construction, since it was asssumed in Section 3.6 to constrain the initial mass accretion rate. On the other hand, similarity between the PMF and the IMF as in Paper 1 is a special case. In general the PMF may vary significantly from the IMF (MO, OM).

### 3.9. Protostar luminosity and luminosity function

The luminosity of an accreting protostar is dominated by its accretion luminosity for a broad range of protostar masses. This property allows the observable accretion luminosity to be used as a probe of the protostar mass, which is not generally observable by other means. The accretion luminosity can be written

$$L = \frac{\gamma G m \dot{m}}{R_\bigstar} \tag{13}$$

where the luminosity efficiency $\gamma$ is the ratio of the accretion luminosity from the star-disk system to that due to spherical accretion onto a spherical surface of radius $R_\bigstar$.

Most models assume that accretion occurs through a circumstellar disk. For steady accretion of infalling material through an optically thick disk, $\gamma = 1/2$ (Hartmann 1998). It is less clear how much kinetic energy of accretion reaches the protostar surface to be radiated as accretion luminosity. Assuming disk accretion and that half the kinetic energy is radiated away before it reaches the protostar, $\gamma = 3/4$ (Offner et al. 2009). Assuming disk accretion and that none of the kinetic energy is radiated away, $\gamma = 1$ (Klassen et al. 2012). These estimates neglect the possibility of nonsteady or episodic accretion, which can further reduce $\gamma$ (Dunham & Vorobyov 2012).

The radius of the accreting protostar also depends on how much energy the accreting mass brings to the protostellar surface. For spherical accretion, the maximum energy is deposited, and the radius increases with both stellar mass and mass accretion rate (Stahler et al. 1986, Hosokawa & Omukai 2009). In contrast, it is expected that accretion of the same mass through a circumstellar disk, or through magnetospheric accretion columns, allows more energy to be radiated away, bringing less energy to the protostar surface.

In the limit of "cold disk accretion" the protostar radius is significantly smaller than for spherical accretion, and it varies less with protostar mass and accretion rate (Palla & Stahler 1992, Hartmann et al. 1997, Hosokawa et al. 2010). These properties of spherical and cold disk accretion apply during the "convection" phase, until the protostar has aquired a mass of several $M_{\odot}$, depending on the accretion rate. Then the "swelling" phase begins, as the surface layers of the protostar rapidly expand (Stahler 1988), in response to the outward transport of entropy (Hosokawa et al. 2010).

Here simplified descriptions of these limiting cases are used to bracket the likely protostar radius, and its corresponding accretion luminosity. It is expected that the most accurate description lies between these limiting cases. In the following expressions, subscripts $s$ and $c$ refer respectively to spherical and cold-disk acccretion.

In the spherically symmetric limit, the radius is approximated by

$$R_{\star s} = R_{0s}\left(\frac{m}{m_1}\right)^{\alpha}\left(\frac{\dot{m}}{\dot{m}_1}\right)^{\beta} \tag{14}$$

where $R_{0s}$= 26 $R_{\odot}$, $m_1$ = 1 $M_{\odot}$, $\dot{m}_1$ = $10^{-3}$ $M_{\odot}\mathrm{yr}^{-1}$, $\alpha$ = 0.27, and $\beta$ = 0.41 as given by Stahler et al. (1986). This relation agrees within a factor ~2 with the stellar models of Hosokawa & Omukai (2009) for masses greater than 0.1 $M_{\odot}$. For the accretion rate in equation (1), equation (14) applies up to a maximum mass ~ 5 $M_{\odot}$, when the radius begins to swell.

In the cold-disk limit, the radius is assumed to have the constant value $R_{\star c} = R_{0c} = 2.5R_{\odot}$. This radius value is typical for low-mass spherical models and it lies below most higher-mass models of cold-disk accretion (Stahler 1988, Hartmann et al. 1997, Hosokawa et al. 2010). It is assumed to apply for masses 0.05 - 5 $M_{\odot}$. However unlike equation (14) it is not an analytic fit to cold-disk accretion models. It should be viewed as a constant-radius model rather than a true cold-disk accretion model.

For spherically symmetric accretion, the accretion luminosity can be written using equations (13) and (14),

$$L_s = L_{0s}\mu^{1-\alpha}\nu^{1-\beta} \tag{15}$$

where the luminosity scale is

$$L_{0s} \equiv \frac{\gamma_s G m_1^{\alpha} \dot{m}_1^{\beta} m_0^{2-\alpha-\beta}}{R_{0s}(\Gamma_s \tau)^{1-\beta}} \quad . \tag{16}$$

In the cold-disk approximation, the luminosity is

$$L_c = L_{0c}\mu\nu \tag{17}$$

where the luminosity scale is

$$L_{0c} \equiv \frac{\gamma_c G m_0^2}{R_{0c}\Gamma_c \tau} \quad . \tag{18}$$

Equations (15)-(18) show that at low mass compared to the mass scale, the luminosity increases with mass to the power 0.71 - 1.0, while at high mass, the luminosity increases more rapidly with mass, to the power 1.3 - 2.0. For the same mass, the cold disk luminosity exceeds the spherical luminosity because the radius is generally greater in the spherically symmetric case.

To evaluate typical luminosities the parameter values $\sigma$ = 0.19 km s$^{-1}$, $\varepsilon$ = 1/3, $\tau$ = 0.12 Myr and $\Gamma_c = \Gamma_s = 1.35$ are adopted, which match the 2CA model to the IMF. The luminosity efficiencies are $\gamma_c$ = 1/2 for cold disk accretion and $\gamma_s$ = 3/4 for spherical accretion, following the above discussion. Then the luminosity scales are $L_{0s} = 6.03\ L_\odot$ and $L_{0c} = 1.88\ L_\odot$. For protostar mass 0.05, 1, and 5 $M_\odot$ equations (15) and (17) give the accretion luminosity as 0.44-2.1 $L_\odot$, 30-60 $L_\odot$, and 250-1500 $L_\odot$. These accretion luminosities exceed the ZAMS luminosity due to nuclear burning for all masses considered.

For each observed PLF, the cold-disk and spherical accretion models are used to estimate the underlying PMF, from

$$\frac{dN}{d\log m} = \frac{dN}{d\log L}\frac{d\log L}{d\log m} \qquad (19)$$

provided the luminosity increases montonically with mass. The term *dlogL/dlogm* is evaluated for each of the limiting luminosity-mass relations (15) and (17).

The relations between accretion luminosity and mass in equations (13) - (19) are based on an accretion history which is one of several which can fit observed luminosity distributions (OM). These relations assume that the mass accretion is a smooth function of time, neglecting fluctuations due to episodic accretion (Dunham & Vorobyov 2012). Thus the present inference of protostar mass from its accretion luminosity is specific to the 2CA model, and it should be considered more useful for a statistical sample than for individual protostars.

## 4. Comparison with observations

### 4.1. Characteristic luminosities

The 2CA model predicts typical luminosities which match those in nearby cluster-forming regions. When equation (19) is used with the model parameters in section 3.8, the resulting luminosity functions have modal values 0.78 $L_\odot$ for cold disk accretion, and 3.8 $L_\odot$ for spherical accretion. These estimates are similar to modal luminosities in nearby star-forming regions, typically 1-3 $L_\odot$ (Evans et al. 2009, Dunham & Vorobyov 2012, Kryukova et al. 2012).

The 2CA model matches typical observed luminosities because it assumes that accretion is inefficient. The inefficiency $\varepsilon$ in the mass transfer, $\varepsilon$ in the rate of mass transfer from core to protostar, and $\gamma$ in the energy transfer from the disk to the protostar are described in Section 3.1. The 2CA model luminosities do not exceed typical observed luminosities by an order of magnitude, in contrast to the simple spherical collapse models used to describe the "luminosity problem" (Kenyon & Hartmann 1995, hereafter KH; MO, OM).

Similarly inefficient departures from simple accretion flow were detailed in a study of the protostar and outflow in the dense core TMC-1 (Terebey et al. 2006). The efficiency of mass transfer from core to protostar was estimated to be 0.6, the efficiency of the mass accretion rate rate from core to protostar was 0.3, and the efficiency of energy transfer from disk to protostar was 0.6, for a combined luminosity efficiency of 0.09 with respect to spherical isothermal collapse onto a spherical protostar. This paper adopts the corresponding factors of 0.3, 0.3, and 0.50 - 0.75 for a combined efficiency of $\gamma\varepsilon^2$ = 0.06-0.09.

The luminosity expression of KH (their equation (2)) assumes perfectly efficient spherical flow onto a spherical protostar surface. Application of the above efficiency estimates to KH equation (2) yields accretion luminosities 0.83 - 2.1 $L_{\odot}$, in good agreement with typical luminosities in the KH sample.

It remains to better understand the physical basis of the inefficiencies assumed in this paper, in terms of the structure and duration of accretion flows and dispersing flows. Furthermore, there is significant evidence for temporal inefficiency of mass transfer to the protostar in the form of episodic or intermittent accretion, due to disk instabilities (KH, Hartmann et al. 1997, Evans et al. 2009, OM, Dunham & Vorobyov 2012). The present model assumes only steady accretion. Thus a complete treatment of the luminosity problem requires better understanding of both the spatial and temporal inefficiencies with respect to idealized models of steady spherical accretion.

## 4.2. Luminosity distributions

Recent observations at mid-infrared and far-infrared wavelengths with the *Spitzer* and *Herschel* telescopes have greatly improved our knowledge of protostar luminosity distributions in nearby star-forming regions (Evans et al. 2009, Dunham et al. 2010, Kryukova et al. 2012). These distributions are referred to here as PLFs (OM). In several regions the PLF has a single peak near 1 $L_{\odot}$ and a high-luminosity tail extending to $10^2$ - $10^3$ $L_{\odot}$ (Evans et al. 2009, Dunham et al. 2010, Kryukova et al. 2012, Kryukova 2011).

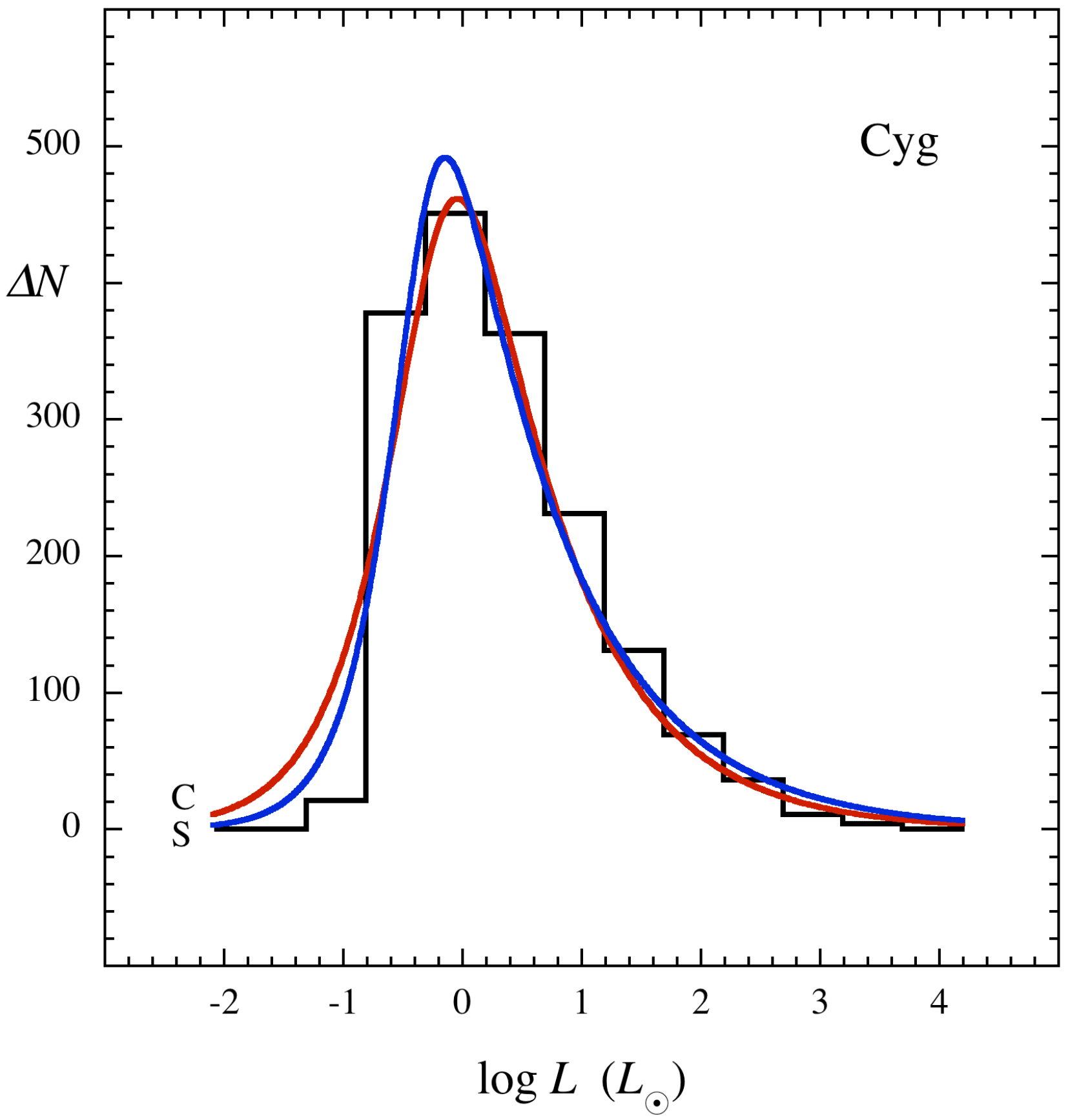

**Figure 2.** Protostar luminosity function for 1695 protostars in Cygnus-X (histogram; Kryukova 2011) and model PLFs for cold disk accretion (red curve, *C*), and spherical accretion (blue curve, *S*).

Observed PLFs have been compared to theoretical predictions based on various accretion models of star formation and on the requirement that the model reproduce the IMF (MO, OM, Paper 1). Here, PLFs are fit in order to compare the underlying mass functions with each other and with the IMF. It is assumed as before that the underlying mass distribution has the form of equation (8) based on the 2CA accretion model and on random accretion stopping. However the mass scale and high-mass slope are now allowed to depart from those which match the IMF. The corresponding model PLF is obtained from equation (19) for each version of the protostar radius. For each cluster, each of these two model PLFs is fit to the observed luminosity distribution. Then each fit PLF is used to infer the underlying mass function.

Observed PLFs and model fits are presented for all of the currently available regions having observed luminosity estimates for at least ~ 100 protostars, in Cygnus (Figure 2; Kryukova 2012), and in Orion A (Figure 3), Mon R2 (Figure 4), and Cep OB3 (Figure 5; Kryukova et al. 2012). A PLF comparison with the "c2d" ensemble drawn from several nearby clouds was given in Dunham et al. (2013).

The PLF fit requires three parameters--the number of protostars, the luminosity scale, and the high-mass slope. The luminosity scale and high-mass slope were chosen so that the model PLF passes through all or nearly all of the histogram bins. This goal was achieved on the high side of the luminosity peak, and one or two bins are missed on the low side, where the data are less complete (Kryukova et al. 2012). The best-fit PLFs for cold-disk and spherical accretion agree well with each other for each cluster. The PLF shape for spherical accretion is slightly more asymmetrical than for cold disk accretion.

Over all four clusters, the luminosity scales $L_{0c}$ and $L_{0s}$ span the narrow range 0.6 - 1.8 $L_{\odot}$. The underlying mass functions have negative log-log slope $\Gamma_c$ = 1.0 -1.8 for cold disk accretion, bracketing the Salpeter slope of the IMF, and significantly shallower slopes $\Gamma_s$ = 0.6 - 0.9 for spherical accretion. Table 2 summarizes these fit parameters.

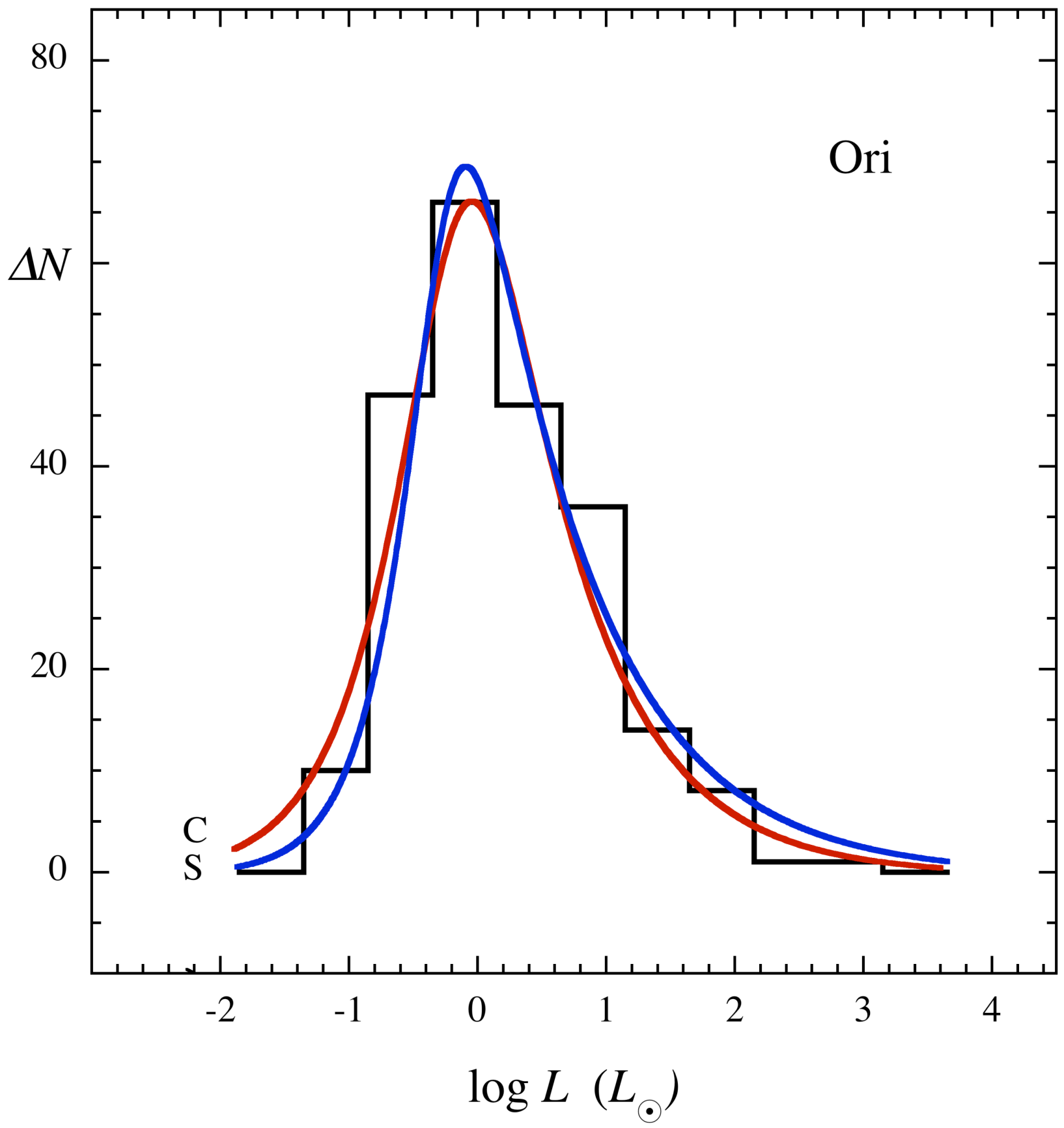

**Figure 3.** Protostar luminosity function for 229 protostars in Orion A (histogram; Kryukova et al. 2012) and model PLFs for cold disk accretion (red curve, *C*), and spherical accretion (blue curve, *S*).

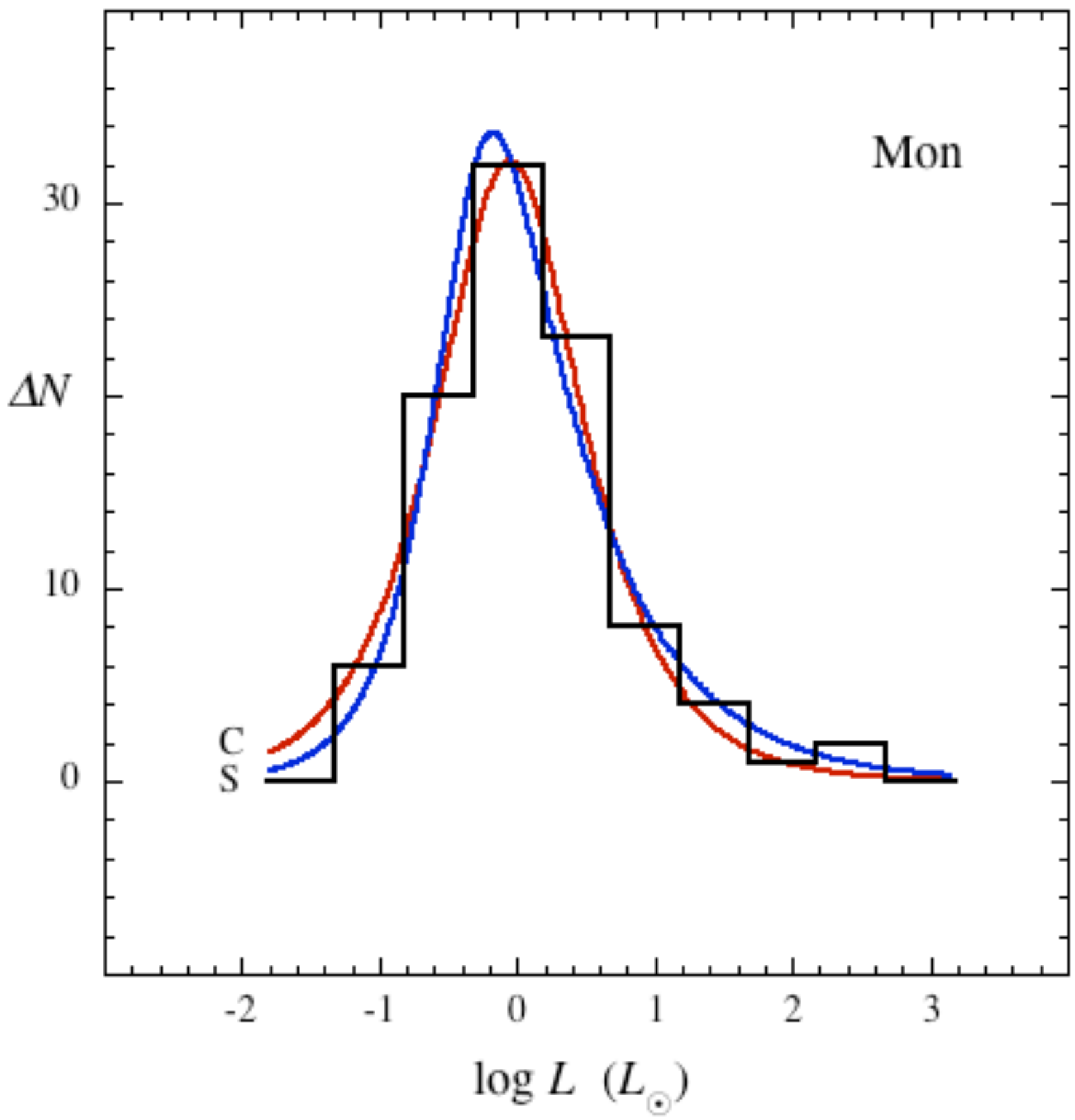

**Figure 4.** Protostar luminosity function for 96 protostars in Mon R2 (histogram; Kryukova et al. 2012) and model PLFs for cold disk accretion (red curve, *C*), and spherical accretion (blue curve, *S*).

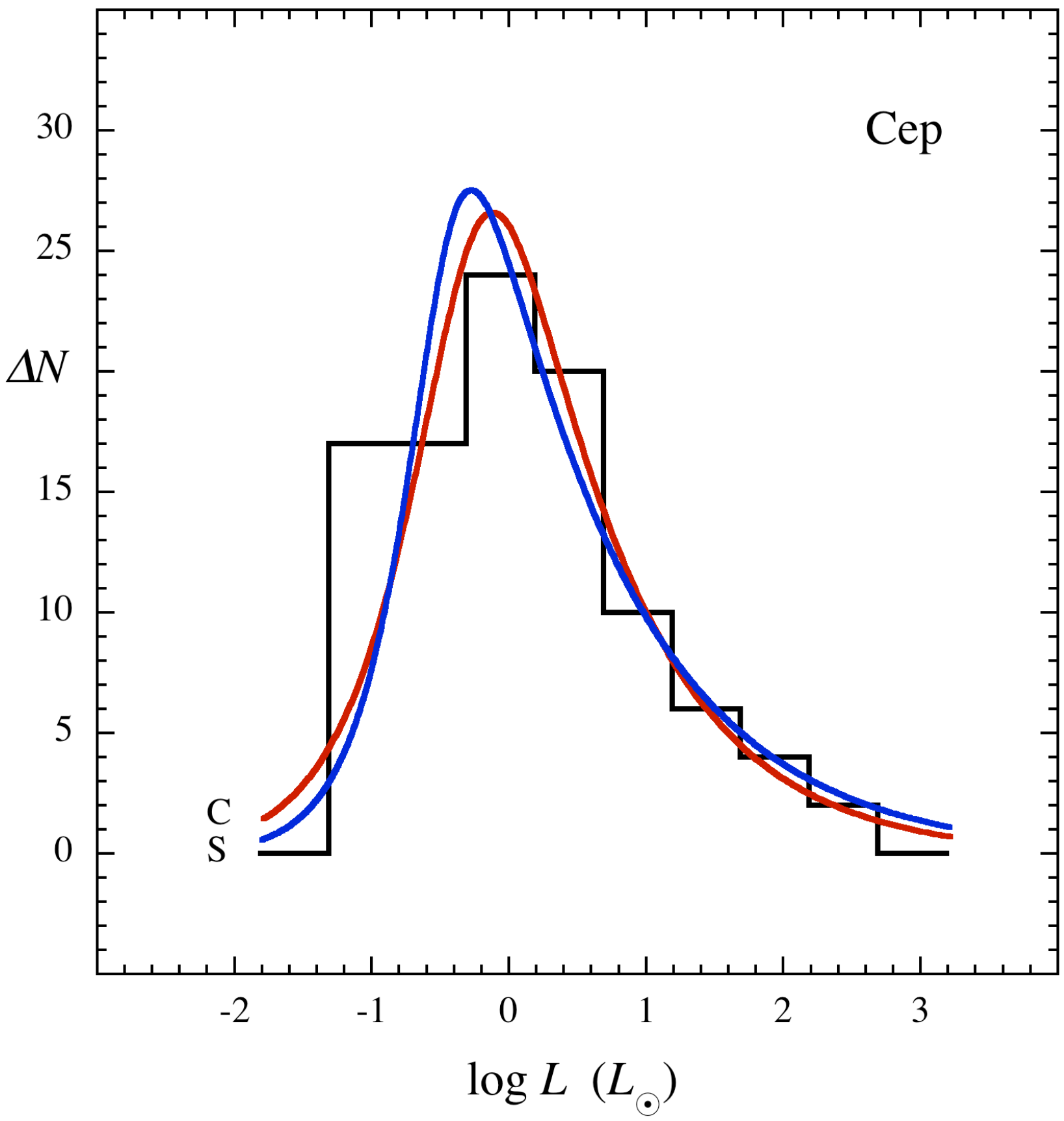

**Figure 5.** Protostar luminosity function for 100 protostars in Cep OB3 (histogram; Kryukova et al. 2012) and model PLFs for cold disk accretion (red curve, *C*), and spherical accretion (blue curve, *S*).

### 4.3. Estimating PMFs from PLFs

The fits of model PLFs to observed PLFs in Section 4.2 suggest that each underlying protostar mass function (PMF) can be inferred by fitting the model luminosity function to the observed PLF. The PMF is useful because it reflects the birth history of cluster stars and can test different theories of star formation (MO, OM). For constant birthrate, 2CA accretion, and random stopping it has the same shape as the final stellar mass function (Paper 1).

However, it is not clear whether the typical embedded cluster satisfies the assumptions of constant birthrate, a steady state between birthrate and accretion stopping, and other conditions of the 2CA model with random stopping. In this section mass functions are inferred from the PLFs fit in Section 4.2, to compare model mass functions with each other and with the IMF.

In estimating a PMF from an observed PLF, it is not required here that the PMF match the IMF. The accretion law and stopping probability each have the same form as in Section 3, so the PMF again tends to a power law at high mass. However the temperature of the mass scale CIS is allowed to depart from the initially assumed 11 K, and the initial mass accretion rate is allowed to depart from the rate of SIS collapse. Thus the mass scales $m_{0s}$ and $m_{0c}$ and the power law exponents $\Gamma_s$ and $\Gamma_c$ are not constrained to the standard IMF values.

To obtain the PMF, the exponents $\Gamma_s$ and $\Gamma_c$ are determined directly from the fits to the PLF. For a PLF with negative log-log slope $\Gamma_L$ at high luminosity, $\Gamma_s = (2-a-\beta)\Gamma_L$ and $\Gamma_c = 2\Gamma_L$, based on equations (8) and (19). To obtain the PMF mass scales, the PLF fit values of $L_{0s}$ and $\Gamma_s$ for spherical accretion are combined with an assumed value of $\gamma_s/\tau^{1-\beta}$ in equation (16) to obtain the mass scale $m_{0s}$. The fit values of $L_{0c}$ and $\Gamma_c$ for cold disk accretion are combined with an assumed value of $\gamma_c/\tau$ in equation (18) to obtain the mass scale $m_{0c}$. The two limiting PMFs are then obtained for each cluster from equation (8).

In each case the star-forming time scale $\tau$ is assumed to be 0.1 Myr as in Section 3.8, and the luminosity efficiencies are $\gamma_c = 0.50$ and $\gamma_s = 0.75$ as discussed above. The mass range is 0.1-5 $M_\odot$ for spherical accretion, limited by the range where equation (14) is valid. The mass range is 0.05-5 $M_\odot$ for cold disk accretion, where the lower minimum value is chosen to reveal the turnover mass of the mass function.

The mass functions are shown for each cluster in Figures 6-9, for cold-disk and spherical accretion. Table 2 gives for each cluster and accretion model, the assumed star-forming time scale and luminosity efficiency; and the inferred mass scale, initial mass accretion rate, and the initial mass scale gas temperature. The mass scale is obtained from equations (15) - (18). The initial accretion rate is obtained from $\dot{m}_0 = m_0/(\Gamma\tau)$ based on equation (7). The kinetic temperature of the mass scale gas is obtained from $T = (m_p/k)[\pi m_0 G/(8\Gamma\varepsilon\tau)]^{2/3}$ based on equation (3), where $m_p = 2.33 m_H$ is the mean particle mass and $k$ is Boltzmann's constant.

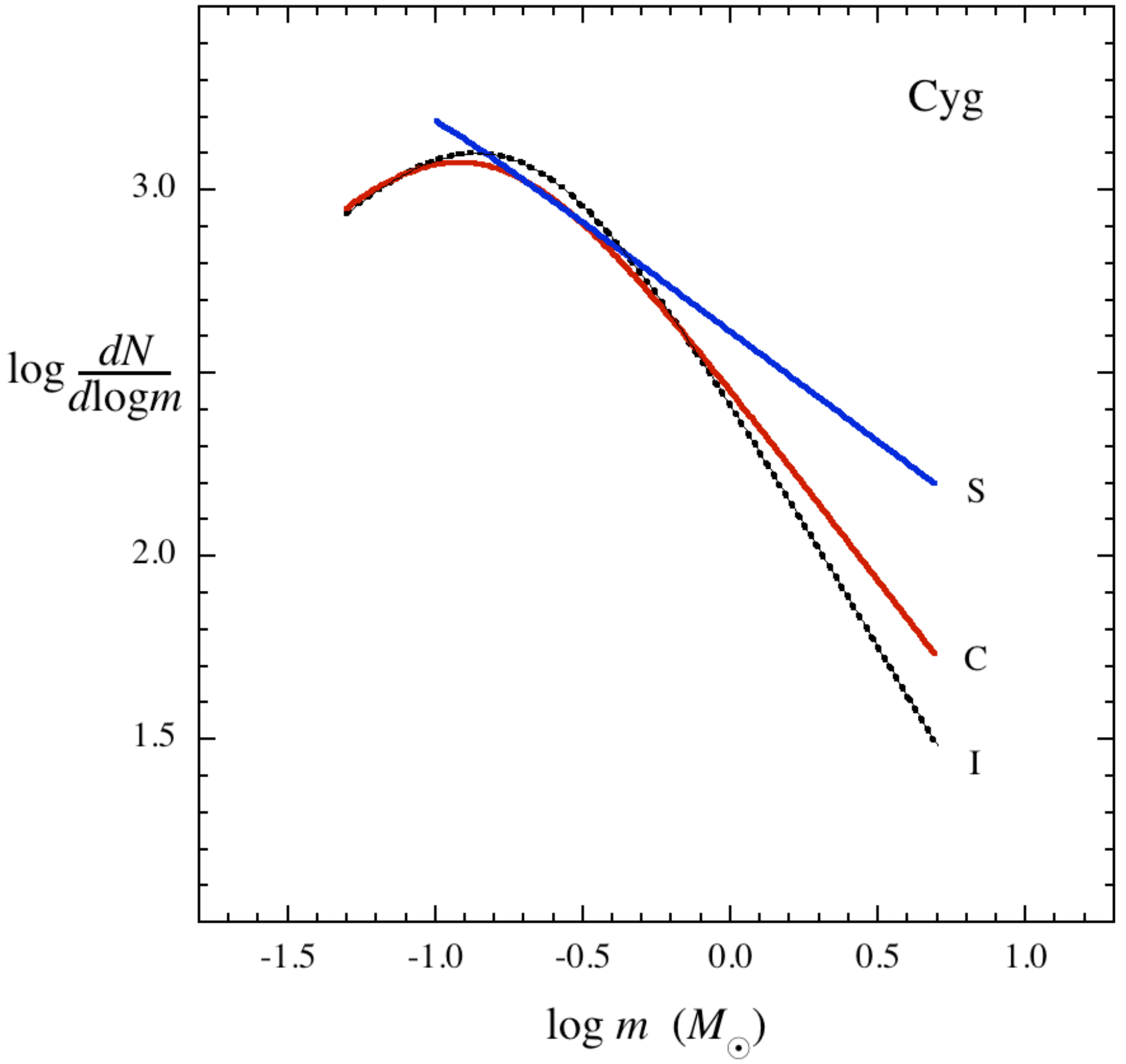

**Figure 6.** Mass functions of 1695 Cygnus X protostars, based on the PLF fits in Figure 2 for spherical accretion (*blue, S*) and for cold-disk accretion (*red, C*), within adopted mass limits. The steady-state mass function which matches the IMF is also shown (*dotted black, I*). The mass function for spherical accretion is displaced upward for clarity.

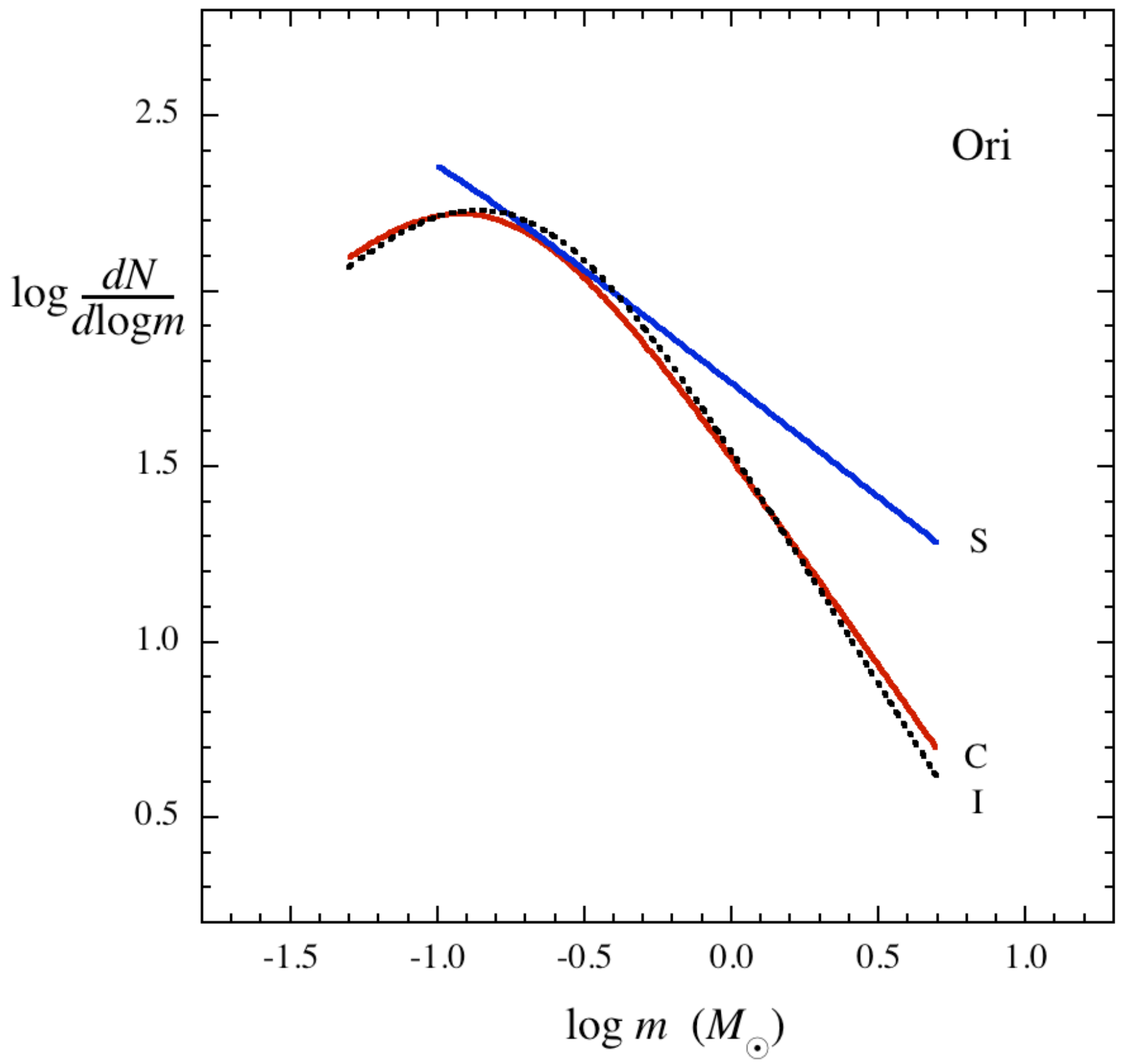

**Figure 7.** Mass functions of 229 Orion A protostars, based on the PLF fits in Figure 3 for spherical accretion (*blue, S*) and for cold-disk accretion (*red, C*), within adopted mass limits. The steady-state mass function which matches the IMF is also shown (*dotted black, I*). The mass function for spherical accretion is displaced upward for clarity.

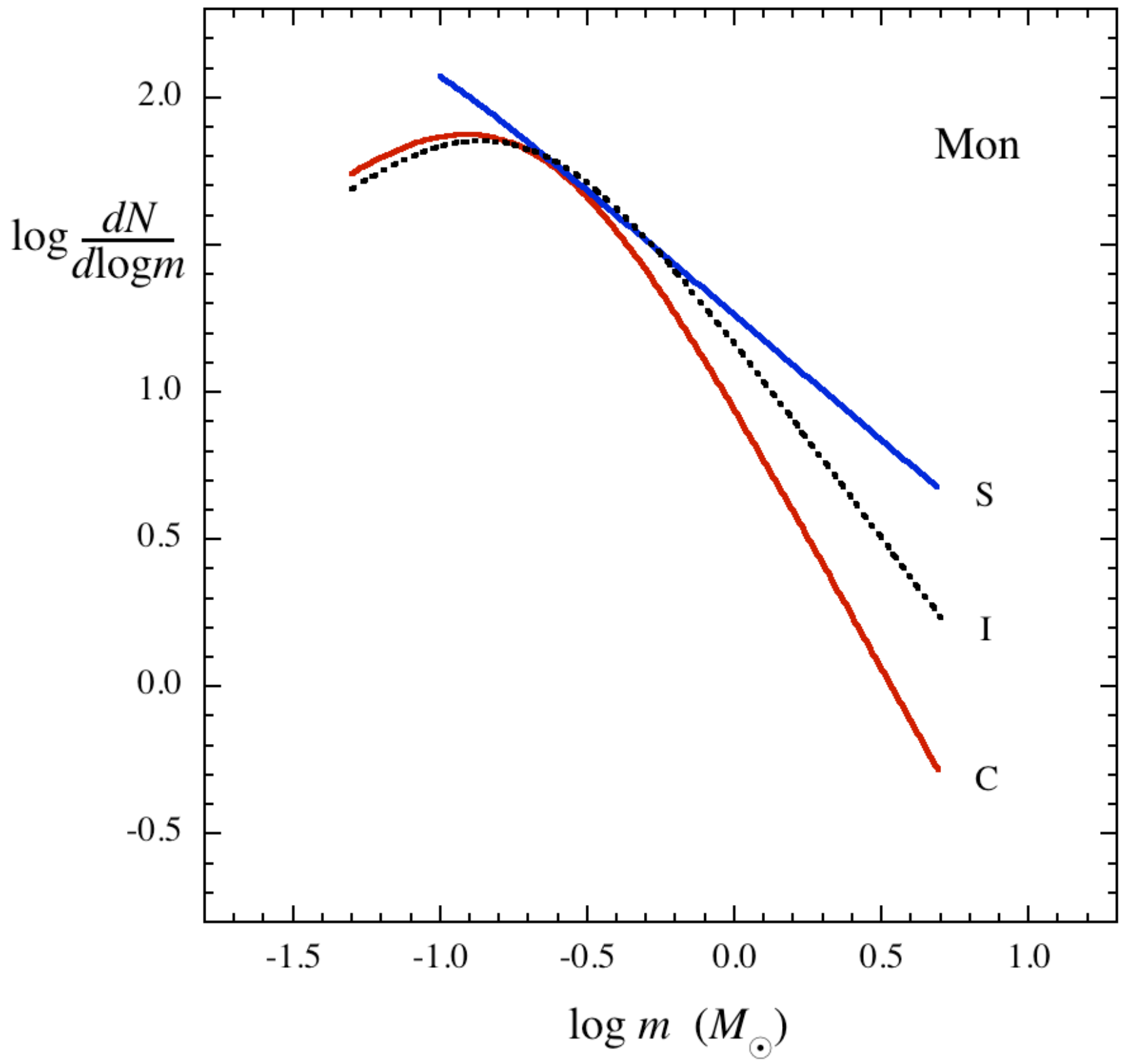

**Figure 8.** Mass functions of 96 Mon R2 protostars, based on the PLF fits in Figure 4 for spherical accretion (*blue, S*) and for cold-disk accretion (*red, C*), within adopted mass limits. The steady-state mass function which matches the IMF is also shown (*dotted black, I*). The mass function for spherical accretion is displaced upward for clarity.

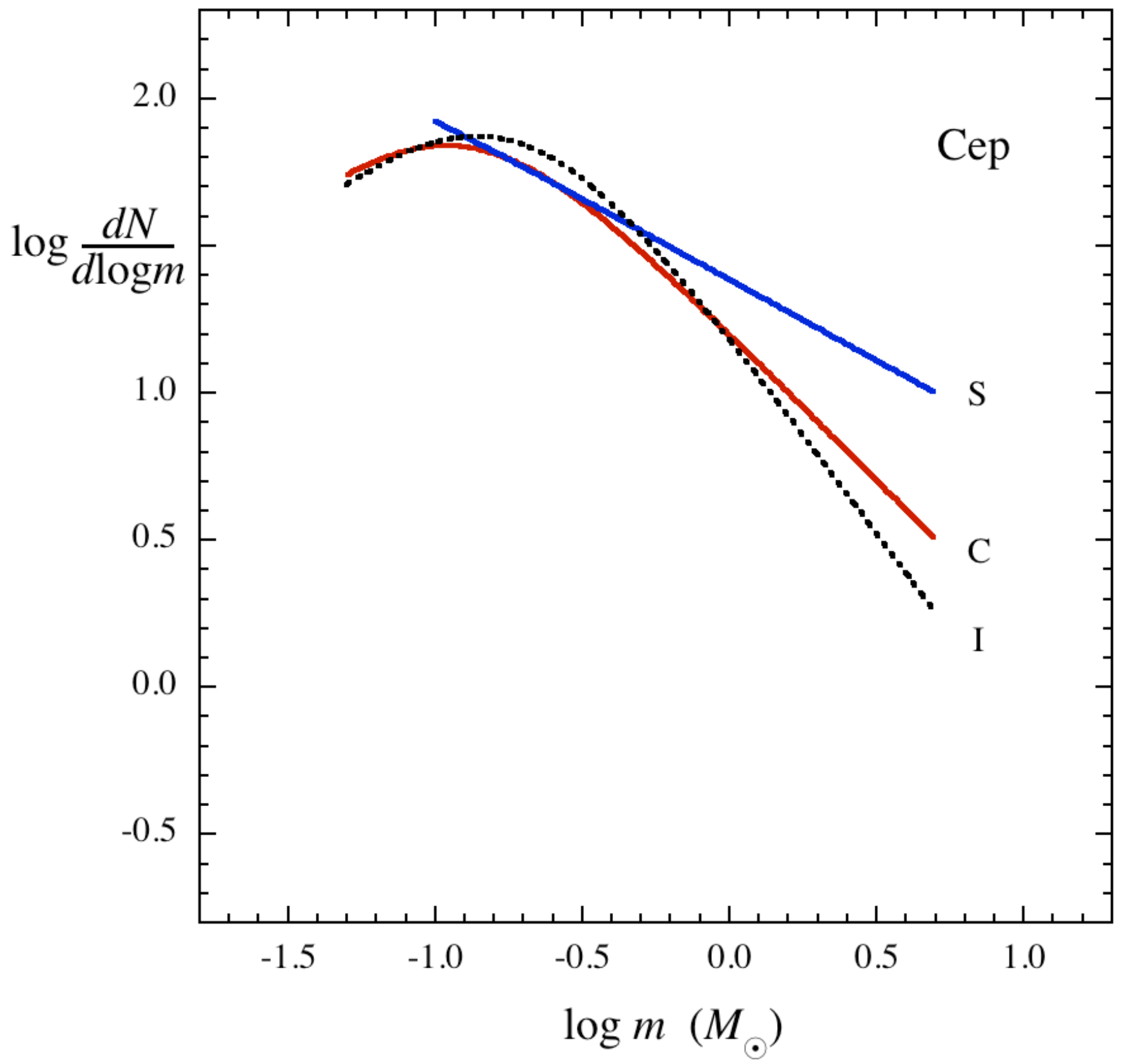

**Figure 9.** Mass functions of 100 Cep OB3 protostars, based on the PLF fits in Figure 5 for spherical accretion (*blue, S*) and for cold-disk accretion (*red, C*), within adopted mass limits. The steady-state mass function which matches the IMF is also shown (*dotted black, I*). The mass function for spherical accretion is displaced upward for clarity.

4.4. PMF mass scales

The protostar mass functions in Figures 6-9 and their properties in Table 2 show that the mass functions for spherical accretion have a mass scale mass lower by a factor ~3 and a high-mass slope shallower by a factor ~ 2 than do the mass functions for cold disk accretion. The mass scale for spherical accretion, typically 0.04 $M_\odot$, causes the modal mass to lie below 0.1 $M_\odot$, the minimum mass where the radius approximation in equation (14) is verified by the stellar structure calculations of Hosokawa & Omukai (2009). In contrast the modal mass for cold-disk accretion lies above its minimum mass. Thus in Figures 6-9 the mass functions for cold-disk accretion show both a high-mass power law and a low-mass turnover, while the mass functions for spherical accretion show only a high-mass power law. Their low-mass turnover is not shown because its location is too uncertain. The mass functions for spherical accretion would require relatively more stars with mass near the peak of the IMF in order to resemble the mass functions for cold-disk accretion.

In contrast to the mass functions for spherical mass accretion, the cold-disk mass functions are relatively similar in mass scale and shape to the IMF. However, evidence for early depletion of Li expected from cold-disk accretion models with high burst accretion rates was not found in a recent search (Sergison et al. 2013).

Table 2 indicates that initial mass accretion rates are 0.4 - 0.5 $M_{\odot}$ Myr$^{-1}$ for spherical accretion and 1.3 - 1.5 $M_{\odot}$ Myr$^{-1}$ for cold disk accretion. Each of these rates is within a factor ~2 of standard estimates for low-mass star formation. However the initial temperatures of mass scale gas are 3-4 K for spherical accretion. These temperatures are significantly less than the 10-15 K values for cold disk accretion, which agree well with $NH_3$ line observations of dense cores (Rosolowsky et al. 2008).

Such low inferred temperatures could increase to match observed values if the product of the time scale $\tau$ and the mass efficiency $\varepsilon$ decreased by a factor ~3. Similarly the mass scale and initial temperature would increase toward observed values if the luminosity efficiency $\gamma_s$ decreased by a similar factor of ~3. However it is difficult to justify any of these changes.

Thus the spherical accretion model of protostar radius appears inconsistent with the models of accretion and accretion stopping, when they are used to fit observed PLFs and to compare with initial gas temperatures. In contrast the constant protostar radius of the cold-disk accretion model is consistent with observed dense core temperatures, but not with calculations of massive protostar evolution, which predict a substantial increase in protostar radius with mass and mass accretion rate (Stahler et al. 1986, Yorke & Bodenheimer 2008, Hosokawa & Omukai 2009).

The simplest resolution of these discrepancies may be that the protostar radius lies between the models of spherical accretion and cold-disk accretion in a way which allows the typical mass scale to be 0.1-0.2 $M_\odot$. This result might arise if low-mass protostars have nearly constant radius, and if the radii of more massive protostars increase less rapidly with mass and mass accretion rate than in equation (14).

## 4.5. PMF slopes

For a PMF inferred from a PLF fit, the slope depends more simply on model parameters than does the mass scale. The slopes inferred for spherical and cold-disk accretion are directly related to the slope of the PLF, as noted in Section 4.3. They have the ratio $\Gamma_s / \Gamma_c = 1 - (\alpha + \beta)/2 = 0.66$ for identical fits to a PLF which declines as a power law at high luminosity. Thus the ratio of PMF slopes obtained from the same PLF differs from unity, mainly because the protostar radius depends differently on mass and accretion rate between spherical accretion, where $\alpha + \beta = 0.68$, and cold-disk accretion, where $\alpha + \beta = 0$. This difference between PMF slopes does not depend on model efficiencies, time scales, or on the choice of initial velocity dispersion.

If the likely radius dependence on mass and accretion rate lies between those of the cold-disk and spherical accretion models, Figures 6-9 and Table 2 suggest that the likely PMF is top-heavy with respect to the IMF. The eight best-fit values of $\Gamma$ in Table 2 have mean ± standard error 1.0 ± 0.1. This typical slope lies below nearly all of the slopes in the compilation of Bastian et al. (2010). The only way the typical inferred PMF could be consistent with the IMF is if the protostar radii generally remain close to the constant value assumed here as a limiting case.

Such top-heavy PMFs are inconsistent with the steady-state cluster model of Section 3, if the mass scale and the initial mass accretion rate are due to thermal equilibrium and collapse, and if the mean accretion duration $t_s$ is equal to $t_f$, the free-fall time of mass scale gas. The top-heavy PMFs cannot be explained by a cluster which is too young to have reached steady state, since then the cluster would have too few stars to be recognized as a large cluster, and it would also have too few massive protostars to match the IMF (Paper 1).

A simple explanation for a top-heavy PMF in a steady state cluster with thermal mass scale and thermal initial accretion rate as in Section 3 is that $t_s$ exceeds $t_f$. If accretion typically lasts longer than the free-fall time of the mass scale gas, the slope of the PMF decreases, as noted in Section 3.6. Such a decrease would be accompanied by an increase in the modal mass, even if the mass scale remained constant. Equation (12) indicates that a decrease in slope from 1.35 to 0.67 would be accompanied by an increase in modal mass by a factor 2.0.

Alternately, a top-heavy PMF might arise because the cluster evolution differs from steady state. First-born stars are expected to experience less stellar feedback from neighbor stars than do later-born stars. Then first-born stars could have average accretion duration which is greater than that of later-born stars, and greater than the free-fall time of the typical core. If so, the slope of the mass function could start out shallow and then steepen as more low-mass stars form. These speculations are discussed further in Section 5.5.

## 5. Discussion

### 5.1. Mass scale

The key features of a mass function model are its mass scale and the high-mass slope. In this model the mass scale is the mass of a CIS, which has the same dependence on mean density and kinetic temperature as the Jeans mass but a smaller coefficient. When equation (3) is expressed in terms of core mean density $n$ and kinetic temperature $T$, the mass scale mass is

$$m_0 = \varepsilon \frac{\Gamma_{IS}}{m_p^2}\left(\frac{6}{\pi n}\right)^{1/2}\left(\frac{kT}{G}\right)^{3/2} \tag{20}$$

where $m_p = 2.33 m_H$ is the mean particle mass, $k$ is Boltzmann's constant, and where $\varepsilon$, $\Gamma_{IS}$, and $G$ have the same values as in Section 3.8.

This mass function mass scale should not vary significantly, if the model is to match the IMF under a wide variety of star-forming conditions. It is difficult to examine this question thoroughly, because the mean density and temperature inferred from observations can vary significantly due to differences in tracer and in resolution. Here the distribution of mass scales is examined for resolution-matched samples of starless cores in low-mass and massive star-forming regions.

The Perseus molecular cloud complex hosts the embedded low-mass clusters NGC1333 and IC348, and a number of more isolated young stars (Bally et al. 2008). The maximum bolometric luminosity of its protostars is 9 $L_\odot$ according to near- and mid-infrared observations (Kryukova et al. 2012). In contrast, a sample of 15 regions within 3.5 kpc of the Sun was selected to reveal conditions for formation of stars of intermediate and high mass in clusters (Sánchez -Monge et al. 2013). Their minimum and median bolometric luminosities are 340 and 3100 $L_\odot$. If the minimum luminosity originates from a single protostar, its mass is at least 3-5 $M_\odot$ according to the models of Section 3.9. Each region has significant clustering, but it is expected that the cluster luminosity is dominated by emission from its most massive members.

The regions were mapped in the $NH_3$ (1,1) and (2,2) molecular lines which trace density and kinetic temperature in star-forming gas (Ho & Townes 1983). These lines were observed with the GBT and the VLA, giving essentially the same linear resolution, 0.04 pc, in 39 starless cores in Perseus (Foster et al. 2009) and in 23 starless cores in the massive star-forming regions (Sánchez-Monge et al. 2013).

Despite the differences in the protostar properties of these samples, the mass scales of their star-forming gas lie within a factor of 2. The number distributions of $m_0$ have first-quartile, median, and third-quartile values of 0.15, 0.17 and 0.19 $M_\odot$ for the low-mass sample, and 0.19, 0.28, and 0.53 $M_\odot$ for the high-mass sample. The mass scale distributions each span a factor ~2 and are marginally distinct. Their widths and separation are much narrower than the mass function itself. Thus the mass function which allows for their spread is negligibly different from a mass function based on their median mass scale. Such a small range in mass scales may arise from the dependence of thermal dust-gas coupling on density (Elmegreen et al. 2008), or possibly from the reduction of fragmentation due to radiative heating in a cluster environment (Bate 2009, Krumholz 2010).

The consistency of mass scales between low-and high-mass star-forming regions within a factor ~2 would not hold if the effective temperature of turbulent motions in the $NH_3$ line widths were used instead of the kinetic temperature. Since the cores in massive star-forming regions have much broader line widths than in low-mass star-forming regions, the median mass scale masses based on line widths differ by a typical factor of ~ 8.

## 5.2. High-mass slope

In this model, the high-mass slope is due to both turbulent and thermal gas properties. The requirement that the mass function approach a power law at high mass is met by the linear dependence of the mass accretion rate on mass, at high mass. This dependence refers to clump gas, which is more turbulent than core gas. However the value of the power law slope, $\Gamma = (m_0 / \dot{m}_0)/\tau$, is due to the mass scale $m_0$ and initial mass accretion rate $\dot{m}_0$, which are set by the physics of the thermally supported core.

It is important to understand more physically how properties of turbulent clump gas lead to the accretion rate needed for the power-law part of the IMF. The dependence of accretion rate on mass found here, $\dot{m} \sim m$, is significantly weaker than for point-mass accretion of nonturbulent polytropic gas, $\dot{m} \sim m^2$ (Bondi 1952). Such a reduction in accretion rate may apply to accretion of gas whose turbulence includes vorticity. An analysis of this process suggests an accretion rate weaker than in pure Bondi accretion, and indicates that inclusion of magnetic fields may further weaken this dependence (Krumholz et al. 2006).

The association of thermal gas properties with the high-mass slope of the IMF might seem to conflict with the observations of massive protostars in gas whose line widths are supersonic (McKee & Tan 2003), and with observations of starless cores with turbulent motions in regions of massive star formation (Wang et al. 2008, Ragan et al. 2012). However in this model, massive protostars are born in thermal cores, and they become massive by accreting both thermal core gas and turbulent clump gas. By the time they are identified as massive due to their high luminosity, they are associated with gas which is turbulent due to its clump origin, and also with core gas which has become more turbulent due to massive star heating, winds, and ionization.

## 5.3. Limitations

It is necessary to specify when accretion stops in order to specify protostar final mass. However the statistical model of equally likely stopping used here is justified mainly by its simplicity. Its mean accretion duration is set by observational estimates, but the form of the distribution itself is neither supported nor refuted by observations and simulations. The likely mechanisms of accretion stopping, including ejection, ionization, winds and outflows, and exhaustion of initial gas have been identified and discussed, but their relative importance over the history of a forming cluster remains unclear.

It is difficult to see how observations can clarify the distribution of accretion durations, but simulations may be helpful, provided they include the relevant physical processes, and can be run long enough to achieve mass functions of final masses. Studies of how simulated protostar masses depend on accretion duration in the presence and absence of outflows, radiative heating, photoionization, and magnetic fields may be useful in developing more realistic distributions of accretion duration.

Furthermore, the same mechanisms which cause accretion to stop are invoked to justify the assumption of a constant efficiency factor $\varepsilon$ between the original gas mass and protostar mass, and between the corresponding mass accretion rates. Although this constant efficiency is a useful simplification, a more serious treatment of dispersal should account for the gradual stopping of accretion, rather than assuming a constant efficiency followed by a sudden stop.

### 5.4. Relation to the core mass function

This model assumes that final protostar masses depend more on differences in accretion duration from one protostar to the next than on differences in initial core properties. Thus the initial accretion rate and the mass scale in equation (1) are each constants, rather than distributions due to core-to-core variations in temperature and mass. Consequently this model is independent of the core mass function, and the resulting mass function is due to accretion and accretion stopping, rather than to core-to-core variations.

Nonetheless it is important to understand the origin of the core mass function, and to understand why the CMF has similar shape to the IMF. Several models of turbulent fragmentation reproduce the CMF (Padoan & Nordlund 2002, Hennebelle & Chabrier 2008, Hennebelle & Chabrier 2009, Hopkins 2012), but these models assume that the IMF masses are a fixed fraction of the corresponding CMF masses, without explaining how this fraction arises. Alternatively it has been suggested that the IMF and the CMF have similar shapes because each is the result of the competition between accretion and dispersal. In one such model, cores grow as filaments contract radially under their self-gravity, and these cores stop growing as outer filament gas is dispersed by stellar feedback, photoionization and evaporation (Myers 2013).

### 5.5. Mass functions in young clusters

As discussed in section 4, the PMFs inferred from the 2CA model and from observed cluster PLFs have typical slope $\Gamma \approx 1.0$, significantly shallower than the IMF.

This result is of interest because it is the first such conclusion based on protostars, having typical accretion age of order 0.1 Myr. In contrast, previous studies of YSO mass functions are based on near-infrared photometry, optical and infrared spectroscopy, and pre-main sequence evolutionary tracks. Such studies of the L1688 cluster in Oph, the Orion Nebula Cluster, the Taurus complex, IC 348, $\sigma$ Ori, $\lambda$ Ori, and Cha I indicate typical YSO ages of a few Myr (da Rio et al. 2012, Bastian et al. 2010, De Marchi et al. 2010).

The top-heavy mass functions in Figures 6-9 are reminiscent of mass functions in the Arches cluster and in NGC 3603. These massive starburst clusters with ages 2-4 Myr show significant mass segregation and flattening of their inner mass function with respect to their outer mass function (Bastian et al. 2010). However as noted above, the mass functions described here refer to much younger stars than those used to obtain mass functions in the Arches and NGC 3603.

A top-heavy PMF might arise from the preferentially early birth of massive stars. Such early birth is a common feature of simulations of cluster formation (Bate et al. 2003, Smith et al. 2009, Wang et al. 2010). In these simulations, the first protostars appear near the center of gravity of their parent clump. Their accretion is relatively long-lived because it is not yet limited by the dispersing effects of winds and outflows from later-born stars. These later-born stars contribute more to the mass function at low mass than at high mass, thereby increasing $\Gamma$ toward the eventual IMF slope. A factor ~2 decrease in accretion time scale over the history of the cluster appears sufficient to account for the shallow slopes in Figures 6-9 and Table 2. However, a more detailed formulation of the accretion stopping probability is needed to test these ideas quantitatively.

## 6. Summary

In this paper a simple analytic model describes protostar mass gain from collapse of a thermally supported core and from accretion of turbulent clump gas. The model is improved over previous versions because its parameters have a more physical basis, its formulation is simpler, and because it is applied in more detail to observed luminosity distributions and cluster mass functions.

The model has these main features:

1. The "2CA" protostar mass accretion rate has a mass-independent component due to thermal core collapse, and a component which depends linearly on mass due to accretion of turbulent clump gas. Protostars stop accreting randomly, due to ejections, stellar feedback, gravitational competition, and exhaustion of initial gas. The mean accretion duration is 0.1 Myr, following estimates of accretion lifetimes in nearby star-forming regions. In a steady state cluster, the combination of mass accretion and random accretion stopping sets the protostar mass function (PMF).

2. Each protostar starts accreting in a thermal core, but its final mass depends on when its accretion stops, not on its parent core mass. Each final protostar mass arises from its original gas mass with efficiency $\varepsilon < 1$. The mass accretion rate of a protostar is similarly inefficient with respect to simple collapse of its original mass. Each of these efficiencies is assumed to equal 1/3, a typical ratio of protostar mass to core mass.

3. The typical original mass is the mass of a critically stable isothermal sphere, which sets the mass scale of the PMF. The initial mass accretion rate is related to the high-mass slope of the PMF. This slope is the ratio of the times scales of accretion and accretion stopping. The gas temperature associated with the mass scale and initial accretion rate is 11 K, following $NH_3$ line observations of starless cores in nearby star-forming regions.

4. The mass scale varies only slightly from low-mass to high-mass star-forming regions, according to observations of starless cores in $NH_3$ lines with similar resolution, ~ 0.04 pc. The mass scale distributions among 39 cores in low-mass regions and among 23 cores in high-mass regions are narrow and similar, having median values 0.17 and 0.28 $M_\odot$ respectively. The model of protostar birth in thermal cores may apply to regions of low-mass and high-mass star formation, despite their differences in protostar luminosity and turbulent motions.

5. The PMF is obtained from the 2CA accretion rate and random stopping, for a steady state cluster where the rate of protostar births equals the rate at which protostars stop accreting, with observed parameter values of temperature, density, and efficiency. The resulting PMF matches the shape of the IMF if the initial mass accretion rate is similar to that of SIS collapse, and if the free-fall time of the typical star-forming core matches the typical duration of accretion.

The model is applied to observations of protostar luminosities in nearby young clusters. The main results are:

1. Accretion luminosities are predicted for protostars which accrete inefficiently and whose PMF matches the IMF. Compared to perfectly efficient models, inefficient accretion reduces the typical luminosity by an order of magnitude factor $(\gamma\epsilon^2)^{-1}$. The predicted typical luminosity is then $\sim 1\ L_\odot$, matching observations.

2. The shape of the protostar luminosity function (PLF) depends on the protostar radius. The radius increases with protostar mass and accretion rate, strongly for spherically symmetric accretion, and weakly for "cold disk" accretion. PLF models based on these limiting cases are used to bracket the likely PMF for a given PLF.

3. Each of the limiting models fits PLFs observed in four clusters with at least 100 protostars, Cyg X, Ori A, Cep OB3, and Mon R2. In these fits, the mass scale and slope of the underlying PMF are allowed to differ from those of the IMF.

4. The PMFs derived for spherical accretion have lower mass scales and shallower slopes than for cold disk accretion. The lower mass scale implies initial gas temperatures which are significantly colder than observed. To match observed gas temperatures, protostar radii should be closer to those of cold disk accretion models than to those of spherical accretion models.

5. The difference in PMF slopes between spherical and cold-disk accretion models depends only on the properties of the protostar radius models. If the slopes derived from the two models are averaged with equal weight, the resulting mean slope is $1.0 \pm 0.1$, indicating a PMF which is “top-heavy” compared to the IMF. Such a top-heavy mass function has been suggested for older, more massive clusters such as the Arches and NGC 3603.

It remains to understand in more detail how the spatial and temporal structure of accretion and dispersal flows account for the inefficiencies which reduce the discrepancy of the luminosity problem. It is speculated that top-heavy PMFs may arise if early-born protostars have less feedback from neighbor stars than later-born protostars. Then early-born protostars would have longer accretion durations and greater final mass than later-born protostars, as is seen in several simulations of cluster formation.

## Acknowledgements

Helpful discussions with Fred Adams, John Bally, Paola Caselli, Rob Gutermuth, Stella Offner, Norm Murray, Ralph Pudritz, Jonathan Tan, and Hans Zinnecker are gratefully acknowledged. Jonathan Foster kindly provided unpublished data associated with Foster et al. (2009). An anonymous referee made numerous suggestions which improved and clarified the paper. Terry Marshall and Irwin Shapiro provided support and encouragement. This material is based upon work supported in part by the National Science Foundation under Grant No. PHY-1066293 and the hospitality of the Aspen Center for Physics.

**Table 1. Variables and Parameters**

| Quantity | meaning or definition | typical value |
|---|---|---|
| $m$ | protostar mass | $0.2\ M_\odot$ |
| $\dot{m}$ | protostar mass accretion rate | $3\ M_\odot\ \mathrm{Myr}^{-1}$ |
| $\dot{m}_0$ | initial $\dot{m}$ | $1\ M_\odot\ \mathrm{Myr}^{-1}$ |
| $m_0$ | protostar mass scale | $0.30\ M_\odot$ |
| $\varepsilon$ | efficiency of accretion and accretion rate | 1/3 |
| $M$ | protostar original mass=$m/\varepsilon$ | $0.6\ M_\odot$ |
| $\dot{M}$ | original mass accretion rate= $\dot{m}/\varepsilon$ | $9\ M_\odot\ \mathrm{Myr}^{-1}$ |
| $M_0$ | original mass scale = $m_0/\varepsilon$ | $0.90\ M_\odot$ |
| $M_{CIS}$ | mass of critically stable isothermal sphere = $M_0$ | $0.90\ M_\odot$ |
| $M_{SIS}$ | mass of truncated singular isothermal sphere having the same temperature and mean density as CIS | $0.67\ M_\odot$ |
| $t_\mathrm{f}$ | free fall time of CIS mean density | 0.1 Myr |
| $\Gamma_\mathrm{IS}$ | isothermal sphere mass ratio $M_{CIS}/M_{SIS}$ | 1.342 |
| $a$ | time since start of protostar accretion | 0.1 Myr |
| $t$ | time since birth of first cluster protostar | 0.5 Myr |
| $t_\mathrm{s}$ | accretion stopping time scale | 0.1 Myr |
| $\mu$ | normalized protostar mass, $m/m_0$ | 1 |
| $\nu$ | $(1+\mu^2)^{1/2}$ | √2 |
| $\Gamma$ | negative slope of mass function | |
| | at high mass, $m_0/(\dot{m}_0 t_s)$ | 1.3 |
| $\Gamma_S$ | Salpeter slope of IMF | 1.35 |
| $\sigma$ | thermal velocity dispersion of | |
| | mass scale gas | 0.2 km $s^{-1}$ |
| *p(a)* | probability density that protostar accretes | |
| | from 0 to *a* and stops between *a* and *a+da* | 5 $Myr^{-1}$ |
| *p(m)* | probability density that protostar accretes | |
| | until its mass is *m* and stops accreting when | |
| | its mass is between *m* and *m+dm* | 5 $M_\odot^{-1}$ |
| $dN/d\log m$ | mass function for *N* objects, ln(10)*Nmp(m)* | 0.7*N* |
| *G* | gravitational constant | $6.67 \times 10^{-8}$ $cm^3$ $g^{-1}$ $s^{-2}$ |
| $\tau$ | star formation time scale, $\tau \equiv t_s = t_f$ | 0.1 Myr |
| $\bar{\mu}$ | normalized mean protostar mass over *p(m)* | 1.7 |
| $\mu_{med}$ | normalized median protostar mass over *p(m)* | 0.5 |
| $\mu_{mod}$ | normalized modal protostar mass over $dN/d\log m$ | 0.6 |
| *L* | protostar accretion luminosity | 1 $L_\odot$ |
| $\gamma$ | efficiency of accretion luminosity compared to | |
| | spherical accretion onto spherical surface | 0.5 |
| $R_\star$ | protostar radius | 2.5 $R_\odot$ |
| $\alpha$ | mass exponent in eq. (14) for $R_\star$ | 0.27 |
| $\beta$ | mass accretion rate exponent in eq. (14) for $R_\star$ | 0.41 |
| $L_0$ | accretion luminosity scale | 2 $L_\odot$ |
| *c* | subscript denoting cold disk accretion | |
| *s* | subscript denoting spherical accretion | |
| $dN/d\log L$ | luminosity function for $N$ protostars, $\ln(10)NLp(L)$ | $0.4N$ |
| $m_p$ | mean particle mass | $3.87 \times 10^{-24}$ g |
| $n$ | mean density of star-forming gas | $3 \times 10^{4}$ cm$^{-3}$ |
| $k$ | Boltzmann's constant | $1.38 \times 10^{-16}$ erg K$^{-1}$ |
| $T$ | kinetic temperature of star-forming gas | 11 K |

Note - variables are listed in the approximate order of their appearance in the paper.

Table 2. Properties of Cluster PLFs and PMFs

| Cluster | $N_{PS}$ | $\Gamma_c$ | $L_{0c}$ | $\gamma_c$ | $\tau_c$ | $m_{0c}$ | $\dot{m}_{0c}$ | $T_c$ | $\Gamma_s$ | $L_{0s}$ | $\gamma_s$ | $\tau_s$ | $m_{0s}$ | $\dot{m}_{0s}$ | $T_s$ |
|---|---|---|---|---|---|---|---|---|---|---|---|---|---|---|---|
| | | | $(L_\odot)$ | | (Myr) | $(M_\odot)$ | $(M_\odot\ \mathrm{Myr}^{-1})$ | (K) | | $(L_\odot)$ | | (Myr) | $(M_\odot)$ | $(M_\odot\ \mathrm{Myr}^{-1})$ | (K) |
| Cyg | 1695 | 1.05 | 1.5 | 0.5 | 0.1 | 0.14 | 1.3 | 10 | 0.60 | 0.78 | 0.75 | 0.1 | 0.030 | 0.49 | 3.6 |
| Ori | 229 | 1.20 | 1.6 | 0.5 | 0.1 | 0.18 | 1.5 | 12 | 0.65 | 0.90 | 0.75 | 0.1 | 0.034 | 0.52 | 4.0 |
| Mon | 96 | 1.80 | 2.1 | 0.5 | 0.1 | 0.25 | 1.4 | 15 | 0.85 | 0.85 | 0.75 | 0.1 | 0.037 | 0.44 | 4.2 |
| Cep | 100 | 1.00 | 1.2 | 0.5 | 0.1 | 0.14 | 1.4 | 10 | 0.55 | 0.55 | 0.75 | 0.1 | 0.022 | 0.40 | 3.0 |

Note - In each cluster, $N_{PS}$ is the number of protostars; $\Gamma$ is the high mass exponent in equation (8) which gives best model fit to the observed PLF; $L_0$ is the best-fit luminosity scale; $\gamma$ is the assumed luminosity efficiency with respect to spherical accretion; $\tau$ is the assumed star formation time scale; $m_0$ is the derived mass function mass scale; $\dot{m}_0$ is the derived initial mass accretion rate; and $T$ is the derived kinetic temperature of mass scale gas. Quantities associated with the protostar radius models for cold disk accretion or spherical accretion are subscripted with *c* or *s* respectively.